**Influential Factors of Users' Trust in the Range Estimation Systems of Battery Electric Vehicles – A Survey Study in China**

**Jiyao Wang**
Intelligent Transportation, Interdisciplinary Programs Office
The Hong Kong University of Science and Technology, Hong Kong SAR, China, 999077
Email: jwanggo@connect.ust.hk

**Chunxi Huang**
Robotics and Autonomous Systems, Interdisciplinary Programs Office
The Hong Kong University of Science and Technology, Hong Kong SAR, China, 999077
Email: tracy.huang@connect.ust.hk

**Ran Tu**
School of Transportation, Southeast University, Nanjing, China, 211189,
Email: turancoolgal@seu.edu.cn

**Dengbo He (Corresponding Author)**
Intelligent Transportation Thrust & Robotics and Autonomous Systems Thrust, Systems Hub
The Hong Kong University of Science and Technology (Guangzhou), Guangdong, China, 511400
Department of Civil and Environmental Engineering
The Hong Kong University of Science and Technology, Hong Kong SAR, China, 999077
Email: dengbohe@ust.hk

Word Count: 5155 words + 3 table (250 words per table) = 5905 words

*Submitted [2022-07-31]*

ABSTRACT

Although the rapid development of battery technology has greatly increased the range of battery electric vehicle (BEV), the range anxiety is still a major concern of BEV users or potential users. Previous work has proposed a framework explaining the influential factors of range anxiety and users' trust toward the range estimation system (RES) of BEV has been identified as a leading factor of range anxiety. The trust in RES may further influence BEV users' charging decisions. However, the formation of trust in RES of BEVs has not yet explored. In this work, a questionnaire has been designed to investigate BEV users' trust in RES and further explore the influential factors of BEV users' charging decision. In total, 152 samples collected from the BEV users in mainland China have been analyzed. The BEV users' gender, driving area, knowledge of BEV or RES, system usability and trust in battery system of smartphones have been identified as influential factors of RES in BEVs, supporting the three-layer framework in automation-related trust (i.e., dispositional trust, situational trust and learned trust). A connection between smartphone charging behaviors and BEV charging behaviors has also been observed. The results from this study can provide insights on the design of RES in BEVs in order to alleviate range anxiety among users. The results can also inform the design of strategies (e.g., advertising, training and in-vehicle HMI design) that can facilitate more rational charging decisions among BEV users.

**INTRODUCTION**

The electrification of automotive power systems is a mainstream solution for current NEVs (new energy vehicles). According to the NEV roadmap proposed by China in 2020, it is estimated that by 2035, China's NEVs will account for more than 50% of total vehicle sales, of which battery electric vehicles (BEV) will account for more than 95% of NEVs (*1*). Compared to fuel vehicles, electric vehicles have several advantages, including fast acceleration and zero emissions. However, range anxiety is one of the main obstacles to the popularization of electric vehicles under current technical conditions (*2*). One of the most widely used definitions of range anxiety is by Rauh et al. (*3*), in which, range anxiety is defined as "*a stressful experience of a present or anticipated range situation, when the range resources and personal resources available to effectively manage the situation (e.g., increase available range) are perceived to be insufficient*". The range anxiety is manifested as the driver's uncertainty about whether they can reach the destination with the vehicle's remaining battery capacity, resulting in cognitive, emotional, behavioral, and physiological changes. Thus, understanding the factors leading to range anxiety can help alleviate BEV users' range anxiety from vehicle design, infrastructure construction and driver training perspectives of view.

In BEVs, the range estimation systems (RESs) can support drivers' recharging decisions before or during the trips. However, the range estimation of the BEV is highly susceptible to the traffic environment (e.g., congestions or not), natural environment (e.g., temperature and weather) and drivers' driving habits (e.g., aggressive or conservative). The difficulty in accurately estimating BEV's range capability may exaggerate drivers' range anxiety. Although the RESs in BEVs have become more and more reliable with the development of technology in recent years, drivers' trust in these systems is questionable. According to the range anxiety model by Rauh et al. (*3*), trust in the RES is one of the dominating factors of range anxiety. However, to the best of knowledge, the factors influencing drivers' trust in the RESs are yet to be explored.

The RES can be regarded as a specific type of automation, which helps users to integrate different parameters of the battery, provide an estimation of the current status of the power system and give energy replenishment suggestions when necessary. Thus, the framework of trust in automation may be adapted to inform the formation of trust in RES. In Hoff et al. (*4*), a three-layer framework has been proposed to conceptualize trust in automation. According to this framework, the trust in automation can be organized into three facets: dispositional trust, situational trust, and learned trust. Each facet can be further influenced by a number of factors. Specifically, dispositional refers to refer to "*long-term tendencies arising from both biological and environmental influences*" (e.g., age and gender of the users); situational trust can be associated with both external variability (i.e., types of the system, its complexity and the difficulty of the tasks) and internal variability (i.e., transitory characteristics that depend on the context); and the learned trust refers to "*an operators' evaluation of the systems learned from past experience or current interactions*" (e.g., knowledge of the system). This framework, according to (*4*), can be used in a variety of scenarios with human operators and automated systems in the loop. However, this framework only provides guidance on potentially influential factors of trust: it may not list all factors that can be related to the RESs in BEVs and conversely, not all factors listed in Hoff et al. (*4*) are relevant to RESs.

In this study, following the trust in automation framework by Hoff et al. (*4*), a questionnaire was designed and distributed to investigate the factors that may influence BEV users' trust in RESs. Further, we investigated how trust in the RESs can affect BEV users' charging behaviors. This study has centered on BEV users in mainland China. In 2021, China contributes 85% percent of EV sales worldwide (*5*). Thus, targeting BEV users in China can provide insights into users' attitudes toward BEVs in a relatively mature market.

## METHODS
### Questionnaire Design

Adopting the framework of trust in automation by Hoff et al. (*4*), the questionnaire explores the potential factors that can influence users' trust in RESs. As this questionnaire was targeted toward BEV users in mainland China, a Chinese version of the questionnaire was used. The translated English version of the questions (Q1 to Q13) is listed in Table 1.

### TABLE 1 Questionnaires

| |
|---|
| Q1: [FI] Date of birth |
| Q2: [SC] Gender at birth |
| Q3: [SC] Please describe the highest level of formal education you have completed |
| Q4: [FI] Please indicate the province you drive the most |
| Q5: [SC] How frequently do your drive BEVs |
| Q6: [FI] The brand of the BEV you drive most frequently |
| Q7: [SC] Please judge the correctness of the following statement<br>- Compared to the slight energy regeneration mode, the one-pedal (or strong energy regeneration) mode reduces the power of BEVs<br>- The battery capacity will be higher at the temperature of 25°C compared to that at 2°C<br>- The BEV with the lithium iron phosphate battery performs better in low temperatures than that of the ternary lithium-iron battery<br>- The range of a BEV is longer at a constant speed of 90 km/h than at 50 km/h<br>- The estimated range will be shorter on a clear city road compared to that on a congested city road<br>- The estimated range will be longer if you always accelerate fast and brake intensively<br>- The time it takes to charge an electric vehicle from 20% to 40% is similar to the time it takes to charge from 80% to 100%<br>- The actual range of a BEV is not related with its power mode (e.g., slight energy regeneration, moderate energy regeneration, and strong energy regeneration) |
| Q8: [LS]The Ten Item Personality Questionnaire (TIPI) (*6*) |
| Q9: [LS] The five-item facets of trustworthiness scale (FIFT) (*7*) regarding users' trust in RES of the BEV they drive the most |
| Q10: [LS] The FIFT for users' trust in State of Charge (SOC) estimation system of the smartphone they use the most |
| Q11: [LS] System Usability Scale (SUS) questionnaire regarding RES of the BEV they drive the most (*8*) |
| Q12: [MC] In what situations do you charge the BEV you use the most |
| Q13: [MC] In what situations do you charge the cellphone you use the most? |

*Note*: Abbreviations of question types are as follow: ***FI***: *Fill-in-text;* ***SC****: Single-choice;* ***MC****: Multiple-choice;* ***LS****: Likert scale;* ***TF****: True or false.*

*Trust in RES of BEVs*

We used a single question out of the five-item facets of trustworthiness scale (FIFT, Q9) (*7*) to assess users' trust in RES of BEV, i.e., "I can trust the range estimation of the BEV." The scale of the question ranges from 1 ("not at all") to 7 ("extremely").

*Dispositional Trust Related Factors*

*Demographic Information*. Research has found that users' trust in automation can vary across different demographic backgrounds, for example, gender (*9*), age (*10*) and education (*7*).

In our questionnaire, we collected information regarding BEV users' age (Q1), gender (Q2), and education (Q3).

*Personality*. According to Hoff et al. (*4*), individuals' personality trait can affect their tendency to trust other people throughout life in the interpersonal domain. Further, personality has been identified as a factor influencing users' comfortable range in BEVs (*12*, *13*). Thus, we adopted the Ten Item Personality Inventory (TIPI) questionnaire (Q8) to measure BEV users' five dimensions of personality, i.e., extraversion, agreeableness, conscientiousness, emotional stability, and openness to experiences. The score for each dimension was calculated following the method proposed in (*6*) and ranged from 1 ("extremely unstable") to 5 ("extremely stable").

*Situational Trust Related Factors*

*Vehicle Brand*. Different functions are provided in BEVs from different manufacturers and the reliability of the RESs also varies across vehicle brands. Further, the research found that people prefer to trust systems that are considered reputable (e.g., *15–17*). Thus, the brand of the BEV was considered a situational-trust related factor and was included in the questionnaire (Q6).

*System Usability*. The ease of system usage plays a significant role in the formation of trust in systems. In the field of e-commerce, numerous studies have found that the high usability of the website is related to high customer trust (*17*, *18*). In the driving domain, Frison et al. (2019) found that the usability of the human-vehicle interface can affect users' trust in automated vehicles (*19*). Thus, we inquired users' impressions on the system usability of their own BEVs. Ten five-point Likert scale (i.e., 1: "strongly disagree" to 5: "strongly agree") items were adapted from the SUS questionnaire (*8*) to assess system usability of RES in BEVs (Q11), in which two facets was assessed, i.e., usability and learnability (both range from 0: "low" to 100: "high").

*Region of Driving*. This factor was assessed by Q4. Because of the significant variance in mainland China's regional nature environment (e.g., temperature) and infrastructure development, we cannot ignore the impact of geographical region on users' attitudes toward the BEVs. In our analysis, we aggregated the provinces in China based on the geographic regional divisions of them, leading to four regions in our analysis, i.e., North, East, South, and Central. It should be noted that, because of the limited samples from west part of China, we aggregated western provinces into North and South regions according to their geographical locations.

*Learned Trust Related Factors*

*Driving Experience*. Previous research found that BEV users' driving style (e.g., driving range, top speed) changes with the hands-on experience with BEVs (*20*). Yuviler et al. (2011) found that users prefer to rely more on the automation system after they gain experience with the automation (*21*), and this may also apply to BEV users. Thus, our questionnaire collected information on drivers' experience with BEVs (Q5).

*System Knowledge*. The knowledge of the BEV and REV might bias users' beliefs in the system's reputation and thus influence their trust in the system (*4*). According to Cohen et al. (*22*), the trust formation process depends on users' level of understanding of how the system would perform in different environments and at different stages in time. Thus, we designed 8 questions assessing BEV users' knowledge of the RESs and BEVs, including the impact of temperature, speed, traffic condition, driving mode and battery type on the performance of the RES, the mechanism of range estimation, and the charging time of batteries (Q7). Participants were expected to give a "true or false" judgement to each statement based on their knowledge and provide confidence ratings of their answers, ranging from 0 ("not sure") to10 ("pretty sure").

For each question, the score was calculated by multiplying the correctness ("correct": 1, "incorrect": -1, and "I don't know": 0) and the confidence rating (possible value ranges from -10 to 10). Finally, the confidence ratings of the 8 questions were weighted and summed to obtain a final score to assess the participants' knowledge (possible score ranges from -80 to 80).

*Experience with Smartphone*. Users' trust in and their experience with the systems similar to BEV (e.g., SOC estimation of the smartphone) may be associated with their trust in RES in BEVs. For example, previous research has identified the transfer of trust between video games and automated combat identification systems (*23*). Considering the wide usage of smartphones and the prevalence of battery anxiety among smartphone users, in our study, we assessed BEV users' experience with and attitudes toward their smartphones, aiming to identify the association between smartphone usage and trust in RESs in BEVs. Users' trust in the SOC estimation of BEV users' frequently-used smartphone was assessed using a single question from FIFT, i.e., "I can trust the SOC estimation of the smartphone" with a Likert scale ranging from 1 ("not at all") to 7 "extremely" (Q10). In addition, Q13 queries users' charging habits of their smartphones, for exploring the potential relationship between users' phone charging behaviors and their charging behaviors of BEVs.

*Users' Behaviors and Attitudes toward BEVs*

Users' trust in the RES may affect users' behaviors when using the BEVs. Thus, we also inquire about BEV users' charging habits of BEVs (Q12).

**Participants**

All participants were recruited via social media on the Internet. A total of 259 participants completed the questionnaire. We then screened the answers based on two quality checking questions (i.e., "please select 4 if you are reading the questionnaire") and 207 samples left after this screening process. Following this, we removed answers from drivers who do not own BEVs or have owned other types of vehicles (e.g., plug-in hybrid electric vehicles or Extended-Range Electric Vehicles) and 164 samples were kept. Next, as commercial vehicle drivers may have developed different strategies in using BEVs and may introduce bias to our samples, we removed the samples from the drivers who are ride-hailing or taxi drivers and finally 152 samples were kept for analyses (Male: 109; Female: 43). These 152 drivers received a compensation of 5 RMB for their completion of the 20-minute-long questionnaire. This study was approved by the Human and Artefacts Research Ethics Committee at the Hong Kong University of Science and Technology (protocol number: HREP-2022-0051).

**Independent Variables**

Table 2 summarizes the definition of independent variables, the levels in the categorical variables, the percentage of samples in each category, and the statistical distribution of each continuous variable (i.e., mean, standard deviation, and minimum and maximum values). For multiple-choice questions (i.e., Q12, Q13), the combination of the categories was treated as levels and only the top combinations were reserved while the rest were aggregated as 'others'.

**Statistics Models**

Two statistical models were built in "SAS OnDemand for Academics", in order to investigate: 1) the influential factors of BEV users' trust in RES (Trust Model); 2) how BEV users' trust, along with other covariates, may affect users' charging behaviors when using BEVs (Charging Behavior Model). Before fitting the models, the correlations between all independent variables were assessed using the Spearman Correlation (*24*) and Pearson Correlation methods (*25*). The highly correlated independent variables were aggregated or abandoned if possible, based on Bayesian Information Criterion (BIC) (*26*) of the models fitted with the variables.

For the Trust Model, a mixed linear regression model was built with PROC MIXED. BEV users' trust in RES (mean:5.6, SD: 0.79, min: 2.5, max: 7) was used as the dependent variable and the independent variables in Table 2, as well as their two-way interactions were used as candidate predictors in the model.

**TABLE 2 Summary of the Independent Variables for Trust Model**

| IV (from Question) | Type | Definition |
|---|---|---|
| Age (Q1) | Continuous | The age of participants measured by years of old<br>- mean: 29.7 (SD: 6.9, min: 18, max: 55) |
| Gender (Q2) | Categorical | - Male (n=109, 71.7%)<br>- Female (n=43, 28.3%) |
| Education (Q3) | Categorical | - Some middle/high schools or less (n=4, 9.2%)<br>- Associate degree (n=31, 20.4%)<br>- Bachelor's degree and above (n=107, 70.4%) |
| Region of driving (Q4) | Categorical | - North (n=64, 42.1%)<br>- Central (n=28, 18.4%)<br>- East (n=32, 21.1%)<br>- South (n=28, 18.4%) |
| Driving frequency of BEVs in past 1 year (Q5) | Categorical | - Frequently (n=137, 90.1%)<br>- Infrequent (n=15, 9.9%) |
| Vehicle brand (Q6) | Categorical | - Tesla (n=63, 41.4%)<br>- BYD (n=37, 24.3%)<br>- Others (n=52, 34.2%) |
| System knowledge (Q7) | Continuous | Participants' knowledge about the BEV and RES<br>- mean: -14.95 (SD: 30.8, min: -60, max: 80) |
| Emotional stability (Q8) | Categorical | - extremely unstable (n= 27, 17.6%)<br>- unstable (n=56, 36.6%)<br>- neutral (n=0, 0%)<br>- stable (n=41, 26.8%)<br>- extremely stable (n=28, 19%) |
| Trust in smartphones (Q10) | Continuous | The score of trust for smartphone<br>- mean: 5.72 (SD: 1.0, min: 2, max: 7) |
| System usability (Q11) | Continuous | The score of system usability<br>- mean: 65.87 (SD: 11.96, min: 40.625, max: 100) |

Note: in this table and the following tables, IV stands for the independent variable, and SD standards for standard deviation.

For the Charging Behavior Model, a multinomial logistic regression model was built using PROC GENMOD. Users' *BEV charging habit* was extracted as the dependent variable from one multiple-choice question regarding their charging behaviors. The candidate independent variables include all independent variables that were abandoned or non-significant (*p*>.1) in the Trust Model, as well as the *Smartphone charging preference* (Q13), which include three levels: charge whenever there is a chance (n=104, 68.4%), charge when battery level is low (n=37, 24.3%) and choose strategy depending on specific situations (n=11, 7.2%). The dependent variable (*BEV charging habit*, Q12) includes four levels: charge whenever a charging station is not far away (n=59, 38.8%); charge if a charging station is at the destination (n=31, 20.4%); choose strategy depending on specific situations (e.g., if a charging station is not far away or a battery warning is sent) (n=29, 19.1%) and others (e.g., charge when the battery capacity is below a user-specify value) (n= 33, 21.7%). It should be noted that, the "*others*" include a number of strategies with too few samples and thus they are aggregated as "others" in our analysis. However, some strategies in "*others*" may worth further analysis if large questionnaire samples are collected.

For both Trust Model and Charging Behavior Model, the full models were built first with all independent variables and their two-way interactions as the predictors and then a backward stepwise selection was performed based on BIC.

**RESULTS**

We report all significant effects ($p<.05$) and marginally significant effects ($.05<p<.1$) in the fitted models after model selection. The post-hoc contrasts were conducted for categorical independent variables if the main or interaction effects were significant or marginally significant. It should be noted that in the model selection process, all dropped variables were not significant ($p>.1$). Table 4 summarizes all statistical results for Trust Model and for Charging Behavior Model.

**TABLE 3 Summary of Model Results**

| Model (DV) | IV (from question # in Table 1) | F-value/$\chi^2$-value | p |
|---|---|---|---|
| Trust Model (Trust in RES of BEVs) | Region of driving (Q4) | F(3, 141) = 3.61 | .01 ** |
| | Gender (Q2) | F(1, 141) = 5.47 | .02 ** |
| | Gender * Region of driving | F(3, 141) = 0.92 | .4 |
| | System usability (Q11) | F(1, 141) = 3.39 | .07 * |
| | System knowledge (Q7) | F(1, 141) = 12.38 | .0006 ** |
| | Trust in smartphones (Q10) | F(1, 141) = 22.57 | <.0001 ** |
| Charging Behavior Model (BEV charging habit) | Smartphone charge preference (Q13) | $\chi^2(6) = 23.61$ | .0006 ** |
| | Driving frequency in past 1 year (Q5) | $\chi^2(3) = 1.97$ | .6 |
| | Education (Q3) | $\chi^2(6) = 6.32$ | .4 |
| | Trust in RES of BEVs (Q9) | $\chi^2(3) = 0.47$ | .9 |
| | Vehicle brand (Q6) | $\chi^2(6) = 1.01$ | .99 |
| | Trust in RES of BEVs * Vehicle brand | $\chi^2(6) = 3.14$ | .8 |
| | Age (Q1) | $\chi^2(3) = 0.83$ | .8 |
| | Age * Vehicle brand | $\chi^2(6) = 6.65$ | .4 |

Note: ** marks significant predictors ($p<.05$) and * marks marginally significant predictors ($.05<p<.1$).

**Trust Model**

Table 3 shows the results of the Trust Model which has an adjusted $R^2$ of 0.282. As shown in Table 3, *Gender*, *Region of driving*, *System knowledge*, and *Trust in smartphones* were found to be significant factors of users' trust in RES in BEVs and the explained variances of these factors were 2.6%, 3.3%, 8.8% and 11.2%, respectively. The System usability was found to have marginally significant effect on trust in RES of BEVs, with an explained variance of 2.2%.

Specifically, for *Region of driving* (as shown in Figure 1a), northern drivers trusted in RES less compared to that of southern drivers, with a difference (Δ) of -0.53 and a 95% confidence interval (CI) of [-0.87, -0.18], t(141) = -3.02, $p$ = .003; northern drivers also trusted less in RES compared to drivers from the central part of China (Δ = -0.40, 95% CI: [-0.75, 0.05], t(141) = -2.24, $p$ = .047). At the same time, a marginally significant difference has been found between eastern and southern drivers, with eastern drivers trusted less in RES compared to southern drivers (Δ = -0.36, 95% CI: [-0.75, 0.04], t(141) = -1.80). Further, for the effect of

*Gender*, as shown in Figure 1b, male drivers trusted less in RES compared to female drivers (Δ = -0.31, 95% CI: [-0.57, -0.05]).

As for the continuous variables, the system knowledge was found to be negatively correlated with the trust in RES of BEVs, while the trust in the SOC of smartphones was found to be positively correlated with the trust in RES of BEVs. Specifically, for each 10-score increase in the *System knowledge*, a 0.07-unit (95% CI: [-0.011, -0.003]) decrease of *Trust in RES of BEVs* has been observed; for each 1 unit increase of *Trust in smartphones*, the *Trust in RES of BEVs* increased 0.3 units (95%CI: [0.16, 0.40]). In addition, a marginally significant effect of *System usability* has been found for *Trust in RES of BEVs*: for each 1 unit increase in *System usability*, the *Trust in RES of BEVs* increased by 0.005 units (95% CI: [-0.0007, 0.0193]).

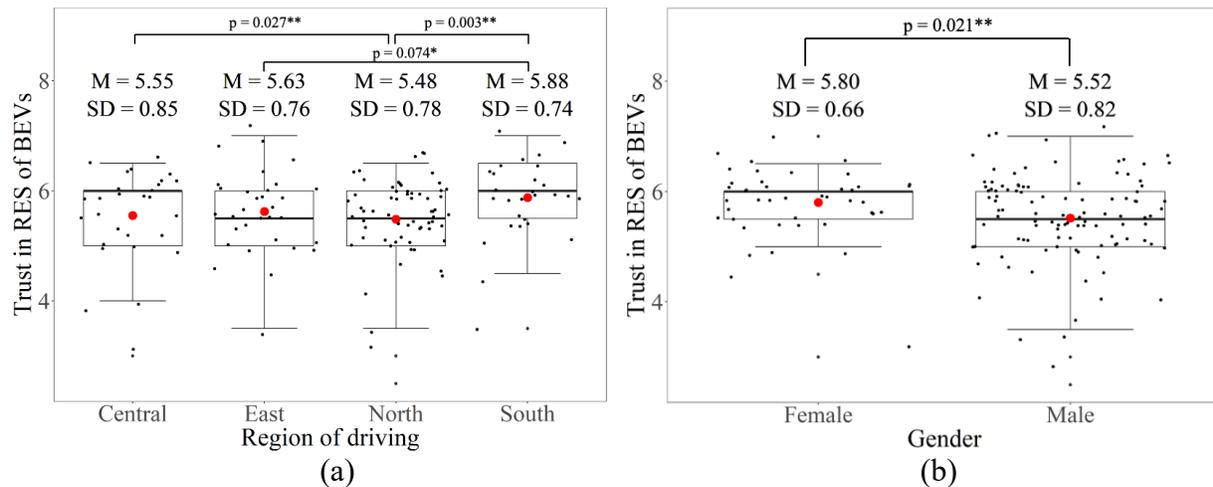

**Figure 1.** *Trust in RES of BEVs* for a) drivers from different regions; b) drivers with different genders. The boxplots present 1.5 times interquartile range below 25th percentile, 25th percentile, median, 75 percentile, and 1.5 times interquartile range above 75th percentile with the bottom whisker, lower edge of the box, bold horizontal line, upper edge of the box, and the top whisker, respectively. Raw data points are indicated with black dots and the averages are indicated with red dots. In the figure, M = mean and SD = standard deviation.

**Charging Behavior Model**

As shown in Table 3, a significant effect of *Smartphone charge preference* has been observed for *BEV charging habit*. Compared to those who "*charge phones depending on specific situations*", those who "*charge phones whenever there is a chance*" are more likely to "*charge their BEVs whenever a charging station is not far away*" relative to "*charge BEVs depending on specific situations*", with an odds ratio (OR) of 12.6 (95%CI: [1.32, 120.2]).

At the same time, compared to those who "*charge phones depending on specific situations*", those who "*charge their phones when the battery level is low*" are also more likely to choose other types of BEV charging strategies ("*others*") that are rare in our samples (OR=16.91, 95%CI: [1.86, 153.81]), more likely to "*charge BEVs whenever a charging station is not far away*" (OR=11.85, 95%CI: [1.32, 106.18]), and also more likely to "*charge BEVs when a charging station is at the destination*" (OR=13.57, 95%CI: [0.97, 190.72]) relative to "*choose BEV charging strategy depending on specific situations*".

**DISCUSSION**

In this study, we adapted the framework of trust in automation by Hoff et al. (*4*) to explore the potential influential factors of trust in RES of BEVs. The results have provided evidence to support the adoption of the framework by Hoff et al. (*4*) to explain the formation of trust in RES of BEVs. In the fitted model, we found that BEV users' trust in RES was associated with all three layers in the trust framework by Hoff et al. (*4*), that is, BEV users' dispositional trust (i.e., Gender), situational trust (i.e., System usability and Region of driving) as well as learned trust (i.e., System knowledge and Trust in smartphones). In our model that fitted the best, the dispositional trust, situational trust and learned trust explained 2.6%, 5.5% and 20% variance of users' trust in RES.

Specifically, gender, as a dispositional-trust-related factor, has been found to influence BEV users' trust toward RES, with females showing higher trust toward RES. The impact of gender on users' attitudes toward automation has been identified in previous research (*9*). It should be noted that, with the questionnaire design in our study, gender is not correlated with any of the independent variables explored in this study. Thus, the difference in trust as a result of gender might be an inherent characteristic of gender and further exploration is needed to explain the underlining factors leading to this difference between different gender groups.

In addition, the region where the drivers drive most often was found to be associated with BEV users' trust in the RES. Northern drivers exhibited the least trust toward RES of BEVs; and eastern drivers exhibited marginally less trust toward RES of BEVs compared to southern drivers, potentially because of the impact of situation complexity and task difficulty on users' level of trust in the system (*27*). Specifically, the BEVs may have different performance in different regions of China due to large differences in environmental conditions across these regions. The performance of the battery is associated with the temperature of the environment; and the range anxiety is closely related to the distribution of the charging facilities (*28*). Both the temperature and the charging facility distribution vary across different regions of China. In general, the northern part of China has a low average temperature that may significantly shorten the available range of BEVs, especially during winter. Both southern and eastern parts of China have a well-developed charging station network, which should alleviate the range anxiety of BEV users, but the southern part of China has a more suitable temperature for BEVs throughout the whole year compared to that of eastern China. These differences in situation difficulty may explain the variance in BEV users' trust toward RES across different regions of China.

At the same time, a positive relationship between *System usability* and *Trust in RES of BEVs* has been observed, indicating a well-designed system can increase users' trust in it (*19*). Further, users' trust in RES was found to be negatively correlated with users' knowledge of the RES and BEV. This result indicates that the more the users know about the RES, the less they trust in it. This might be because those who are less aware of the limitations of a system are more likely to have an inappropriate mental model of the system and thus "over-trust" it. Such a phenomenon has been observed among users of driving automation (*22*). However, the readers should be aware that we were not able to gauge whether a trust in RES was appropriate or not. Future research is needed to explore the appropriateness of trust in RES of BEVs.

Another interesting finding from our analysis is that we have identified a connection between trust in smartphone SOC estimation and trust in RES of BEVs. Both smartphone and RES are equipped with SOC estimation system and users' experience in them may be mutually transferable. However, it should be noted that the correlation is not equal to causality. In other words, users who trust more in SOC estimation in smartphones might be those who are prone to show high trust in automation, and it is not surprising they would show high trust in RES of

BEVs. An experiment with controlled variables (e.g., the reliability of the SOC estimation system of a smartphone) is still needed to confirm whether the experience with smartphone SOC estimation would affect users' trust in RES of BEVs.

The connection between similar systems has also been observed in terms of users' energy replenishment behaviors. Trust in RES of BEVs, surprisingly, did not influence BEV users' charging behaviors but we have observed a connection between users' phone charging behavior and BEV charging behavior. The transferring of habits in using similar systems has been observed in previous research (29) but it is the first time being observed among BEV users. Again, we are not able to decide whether there is a causality between the habits or there are other underlying factors deciding the charging habits both when the users were using BEV and when they were using smartphones. Future observational study or experiments with controlled variables are needed to better reveal the relationship that has been observed in our study.

## CONCLUSION

Low trust in range estimation system (RES) of battery electric vehicles (BEV) has been found to increase drivers' range anxiety. In this study, through an online survey, we investigated the factors that can influence BEV users' trust in RES, and further explored how trust in RES as well as BEV users' experience with other electronic devices can influence BEV users' charging behaviors. The results show that the dispositional-, situational-, and learned-trust related factors all affect users' trust in RES of BEVs, supporting the validity of the three-layer trust framework (4) among BEV users. We have also observed the transferability of users' energy replenishment behaviors when using similar systems, i.e., between smartphones and BEVs. Given the relatively small variance explained by our trust model, future research should explore more potential trust-related factors based on a larger number of samples.

## ACKNOWLEDGEMENT

This research is supported by the start-up funding of the Hong Kong University of Science and Technology (Guangzhou). The authors would like to thank Yi Liu and Wenrui Ye for their help in designing the questionnaire and collecting the data.

## AUTHOR CONTRIBUTIONS

The authors confirm contribution to the paper as follows: study conception and questionnaire design: J. Wang, D. He, R. Tu.; data processing: J. Wang, C. Huang, R. Tu, D. He; analysis and interpretation of results: J. Wang, C. Huang, R. Tu, D. He; draft manuscript preparation: J. Wang, C. Huang, D. He; manuscript revision: R. Tu, D He. All authors reviewed the results and approved the final version of the manuscript.